# Using Intelligent Control to Improve PV Systems Efficiency

Dr. Ibrahim Ighneiwa
Department of Electrical and Electronics Engineering
Faculty of Engineering, University, of Benghazi
Benghazi, Libya
ibrahim.ighneiwa@uob.edu.ly

Abdelmeneim Abduljalil Yousuf
Department of Energy Technology
The Libyan Academy
Benghazi, Libya
rehab201475@gmail.com

*Abstract*—One of the important technologies in renewable energy is the Photovoltaic (PV), which is the direct conversion of light into electricity at the atomic level, and improving the efficiency of PV systems is one of the very important factors in getting the most out of this invaluable renewable resource of energy. While most research work we studied used conventional techniques to control two parameters at most, like power and change in power, or voltage and change in voltage, etc., we implemented unconventional techniques, namely intelligent control to control more than two parameters at a time, including change in temperature which had been ignored by many researchers for various reasons, as well as the use of probability theory to predict the location of power point and control how it would move before not after it did. Practically, we utilized available PV systems devices to test our controlled systems and we used simulations and compare our findings with previous work done by others in this area and our techniques showed good improvement in efficiency and we believe that it could open the door for other colleagues to add valuable work in this important field.

*Keywords—PhotoVoltaic, Intelligent Control, Fuzzy Logic, ProbabilityTheory*

## I. INTRODUCTION

Renewable energy is energy that is collected from renewable resources, which are naturally available such as sunlight, wind, rain, tides, waves, and geothermal heat. Renewable energy often provides energy in many important areas, such as electricity generation, air and water heating and cooling and rural (off-grid) energy [1]. Based on REN21's 2016 report [1], renewables contributed 19.2% to humans' global energy consumption (Fig. 1) and 23.7% to their generation of electricity (Fig. 2).

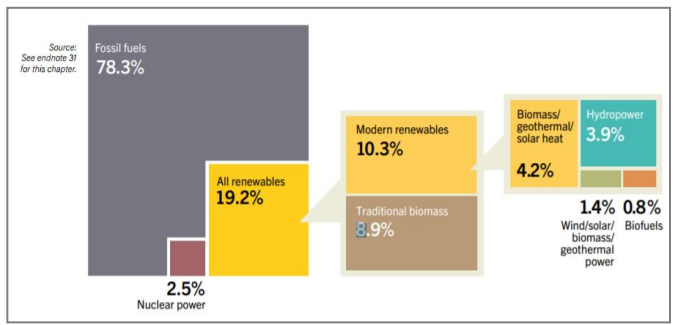

Fig. 1. Renewables contributed 19.2% to global energy consumption

One of the easiest to implement and clean technologies in renewable energy is the PhotoVoltaic (PV), which comprises 1.2% of renewables (Fig. 2) and improving the efficiency of PV systems is one of the very important factors in getting the most out of this invaluable renewable resource of energy. While most research work we studied used conventional techniques to control two parameters, we implemented unconventional techniques (intelligent control) to control more than two parameters at a time.

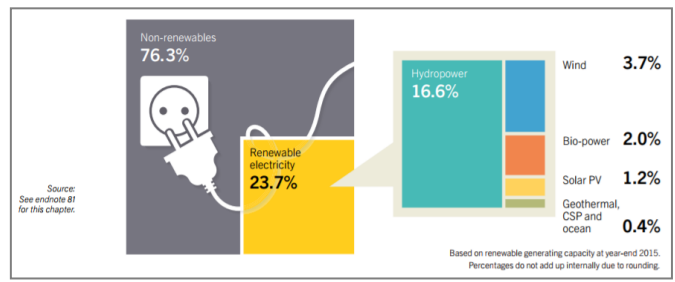

Fig. 2. Renewables contributed 23.7% to global generation of electricity

### A. PhotoVoltaic (PV)

PhotoVoltaic (PV) is the direct conversion of light into electricity at the atomic level. Some materials exhibit a property known as the photoelectric effect that causes them to absorb photons of light and release electrons. When these free electrons are captured, an electric current results that can be used as electricity (Fig. 3) [2].

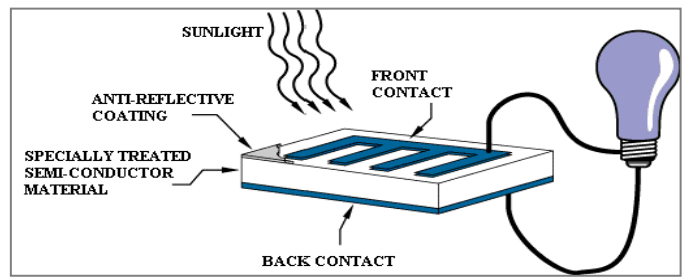

Fig. 3. Photoelectric effect: Materials absorb light and release electrons

PV cells can be modeled as a current source in parallel with a diode. and the single diode as in (1) assumes a constant value for the ideality factor n.

$$J = J_L - J_{01}\left\{exp\left[\frac{q(V+JR_s)}{kT}\right] - 1\right\} - \frac{V+JR_s}{R_{shunt}} \quad (1)$$

In reality the ideality factor is a function of voltage across the device. At high voltage, When the recombination in the device is dominated by the surfaces and the bulk regions the ideality factor is close to one. However at lower voltages, recombination in the junction dominates and the ideality factor approaches 2. The junction recombination is modeled by adding a second diode in parallel with the first and setting the ideality factor typically to 2 (Fig. 4) [3] and (1) must be modified to (2).

$$J = J_L - J_{01}\left\{exp\left[\frac{q(V+JR_s)}{kT}\right] - 1\right\} - J_{02}\left\{exp\left[\frac{q(V+JR_s)}{kT}\right] - 1\right\} - \frac{V+JR_s}{R_{shunt}} \quad (2)$$

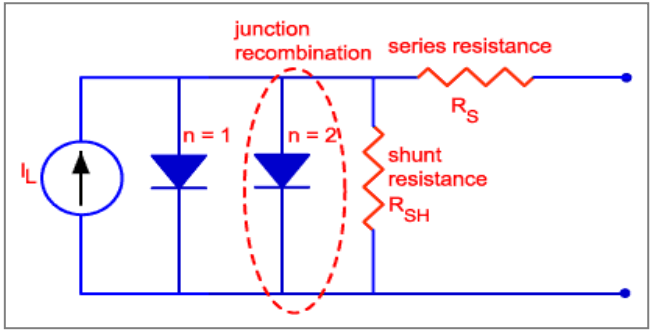

Fig. 4. PV cells can be modeled as a current source in parallel with one diode (n=1) or two diodes (n=2)

When there is no light present to generate any current, the PV cell behaves like a diode. As the intensity of incident light increases, current is generated by the PV cell.

*B. PV Efficiency (η)*

PV Efficiency is the ratio of the electrical power output Pout, compared to the solar power input, Pin, into the PV cell. Pout can be taken to be $P_{MAX}$ since the solar cell can be operated up to its maximum power output to get the maximum efficiency. Fig. 5 shows PV efficiency improvement in the past, present and future, which is still very small.

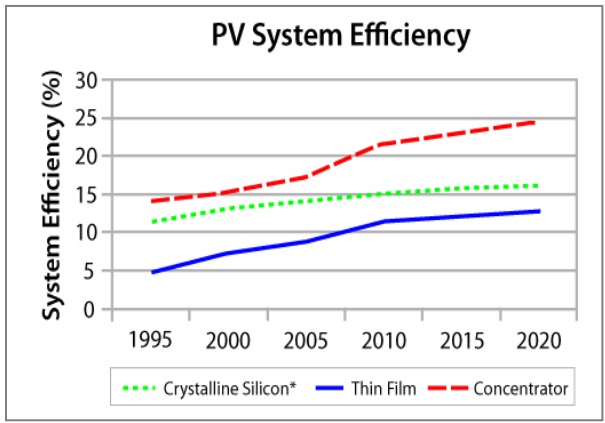

Fig. 5. PV system's efficiency, the past, present and the future

Improving the efficiency of PV is one the very important factors in PV technology, and since we are using an Intelligent Control (IC), namely Fuzzy Logic Control (FLC) to improve PV efficiency; the following two sections will introduce IC and FLC in specific.

*C. Intelligent Control*

In developing a Conventional Control (CC) system to control a plant, the designer constructs a mathematical model of the system. This model contains all the dynamics of the plant that affects controlling it. This type of control is called the Mathematicians Approach, since the designer must mathematically model the plant to be controlled [4]. In developing an Intelligent Control (IC) system to control an agent or plant, the designer inputs the system behaviour and the IC system abstractly models the system. [4].

The types of Intelligent Control includes: fuzzy logic, artificial neural networks, genetic programming, support vector machines, and reinforcement learning, among many others, and since we are using fuzzy logic to improve PV efficiency, we explain this technique in the following section.

*D. Fuzzy logic*

Fuzzy logic is a form of many-valued logic in which the truth values of variables may be any real number between 0 and 1. By contrast, in Boolean logic, the truth values of variables may only be the integer values 0 or 1 (cold or hot). Fuzzy logic has been employed to handle the concept of partial truth (cold, cool, normal, warm, and hot) (Fig. 6), where the truth value may range between completely true and completely false [5].

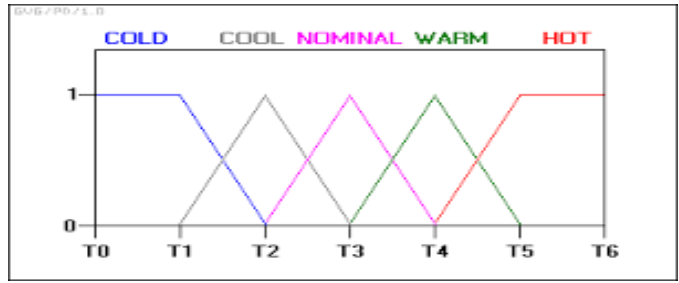

Fig. 6. Fuzzy logic handles the concept of partial truth

In the fuzzy control process, the Fuzzy Inference System (FIS)(Fig. 7) follows the following steps in order to process the crisp input get the final crisp output:
1) Fuzzify input values into fuzzy membership functions.
2) Execute all applicable rules in the rulebase to compute the fuzzy output functions.
3) Defuzzify output functions to get 'crisp' output values.

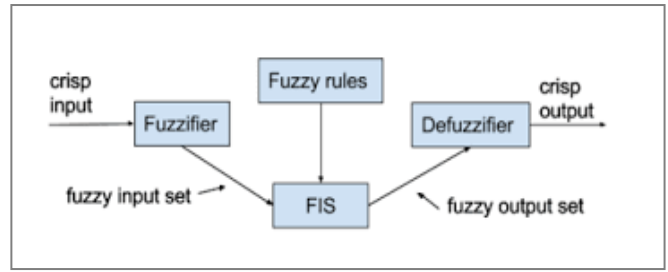

Fig. 7. The Fuzzy Inference System (FIS)

In section II, we review some previous work on improving PV systems efficiency, and in section III we discuss the PV efficiency problem and the PV systems theory. Our utilization of intelligent control to improve PV systems efficiency is detailed in section IV. Finally in section V, we conclude by summarizing our work and suggest some recommendations and suggestions that may help in much more improvement in the PV systems technology.

## II. LITERATURE REVIEW

One of the main problems in PhotoVoltaic (PV) technology is its low efficiency due to many factors, mainly the semiconductor material and the ambient conditions, so controlling the Maximum Power Point (MPP) and tracking it (MPPT) has attracted extensive research in recent years and many new techniques have been reported.

Millet et al. in [6] provides a comprehensive review of the maximum power point tracking (MPPT) techniques applied to photoVoltaic (PV) power system, while reporting that confusion lies while selecting a MPPT as every technique has its own merits and demerits.

In [7], B. Amrouche et al. have shown that the negative effects associated to the conventional control methods such as perturb and observe (P&O) can be greatly reduced if the Artificial Intelligence (AI) concepts are used, where the perturbation step is continuously approximated by using artificial neural network (ANN).

Sedaghati et al. [8] used artificial neural network (ANN) for tracking of maximum power point. They implemented error back propagation method in order to train the neural network.

In this method neural network is used to specify the reference voltage of maximum power point under different atmospheric conditions.

In [9] Kulaksiz and Aydoğdu implemented A maximum power point tracking (MPPT) algorithm using fuzzy controller.

MPPT method was implemented based on the voltage and reference PV voltage value was obtained from Artificial Neural Network (ANN) model of PV modules.

Makhloufi et al. in [10] presented a performance comparison between artificial neural network (ANN) controller and Perturb and Observe (P&O) method has been carried out which has shown the effectiveness of artificial neural networks controller to draw much energy and fast response against change in working conditions.

## III. PHOTOVOLTAIC (PV) AND PHOTOELECTRIC EFFECT

The photoelectric effect was first noted by a French physicist, Edmund Bequerel, in 1839, who found that certain materials would produce small amounts of electric current when exposed to light.

In 1905, Albert Einstein described the nature of light and the photoelectric effect on which photovoltaic technology is based, for which he later won a Nobel prize in physics [2].

The first photovoltaic module was built by Bell Laboratories in 1954. It was billed as a solar battery and was mostly just a curiosity as it was too expensive to gain widespread use. In the 1960s, the space industry began to make the first serious use of the technology to provide power aboard spacecraft [2].

Fig. 8 illustrates the operation of a basic PV cell system, composed of a solar cells, voltage regulators, converters. inverters ..etc. Solar cells are usually made of the some kind of semiconductor materials, such as silicon.

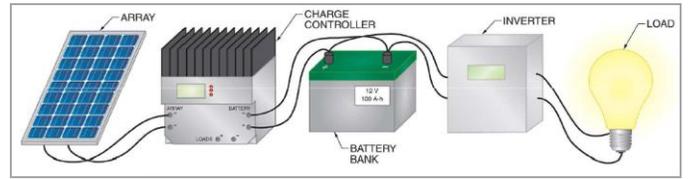

Fig. 8. Basic photovoltaic system

In solar cells, a thin semiconductor wafer is specially treated to form an electric field, positive on one side and negative on the other. When light energy strikes the solar cell, electrons are knocked loose from the atoms in the semiconductor material. A number of solar cells electrically connected to each other and mounted in a support structure or frame is called a photovoltaic module Modules are designed to supply electricity at a certain voltage, such as a common 12 volts system. Multiple modules can be wired together to form an array (Fig. 9) [2].

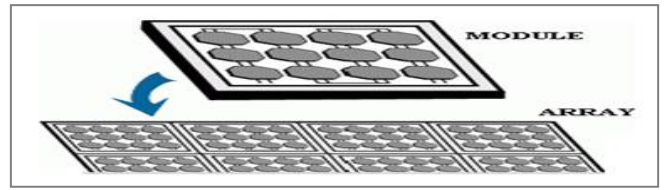

Fig. 9. PV Cells form a module and modules form an array

### A. PV Theory and I-V Characterization

PV cells can be modeled as a current source in parallel with a diode. When there is no light present to generate any current, the PV cell behaves like a diode. As the intensity of incident light increases, current is generated by the PV cell, as illustrated in Fig. 10 [3].

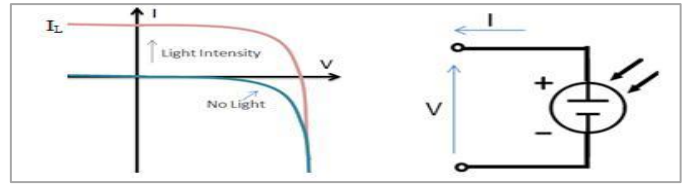

Fig. 10. I-V curve of PV cell and its associated electrical diagram

In an ideal cell, the total current I is equal to the current $I_\ell$ generated by the photoelectric effect minus the diode current ID, according to (3) as follows:

$$I = I_l - I_D = I_l - I_o \left(e^{\frac{qV}{kT}} - 1\right) \qquad (3)$$

where $I_0$ is the saturation current, q is the electronic charge ($1.6 \times 10^{-19}$ Coulombs), k is Boltzman constant ($1.38 \times 10^{-23}$ J/K), T is the cell temperature in Kelvin, and V is the measured cell voltage. Expanding (3) gives (4).

$$I = I_l - I_0 \left(exp^{\frac{q(V+I.R_S)}{n.k.T}} - 1\right) - \frac{V + I.R_S}{R_{SH}} \qquad (4)$$

where n is the diode ideality factor (typically between 1 and 2), $R_S$ contributes to voltage loss and $R_{SH}$ contributes to current loss due to ambient conditions [3].

The I-V curve of an illuminated PV cell has the shape shown in Fig. 11 as the voltage across the measuring load is swept from zero to $V_{OC}$, and many performance parameters for the cell can be determined from this data.

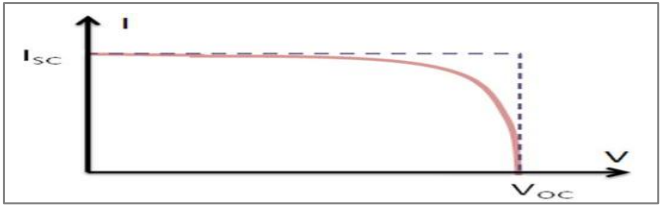

Fig. 11. Illuminated I-V Sweep Curve

The short circuit current ISC corresponds to the short circuit condition when the impedance is low and is calculated when the voltage equals to 0. The open circuit voltage ($V_{OC}$) occurs when there is no current passing through the cell.

Considering different temperatures and solar irradiations [11], the simulated output characteristics of the PV array are depicted in Fig. 12 through Fig 15.

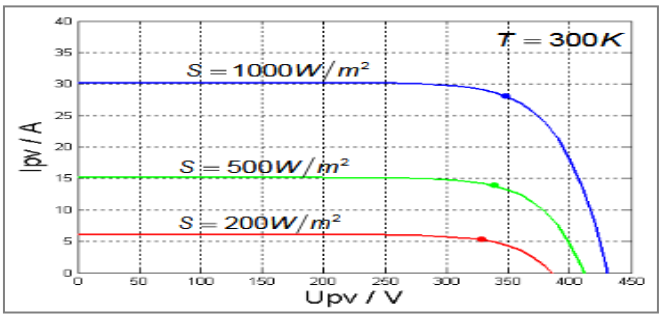

Fig. 12. I vs. V as irradiance changes at T=300K

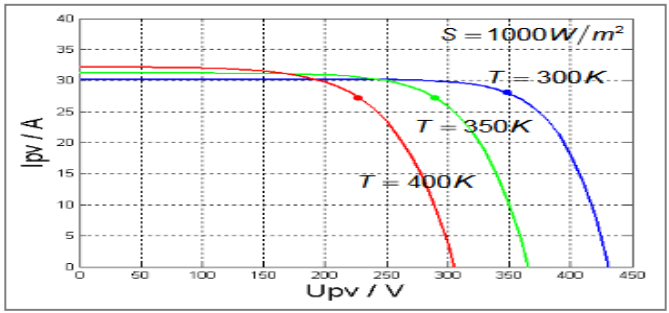

Fig. 13. I vs. V as temperature changes at irradiance (S) =1000W/m2

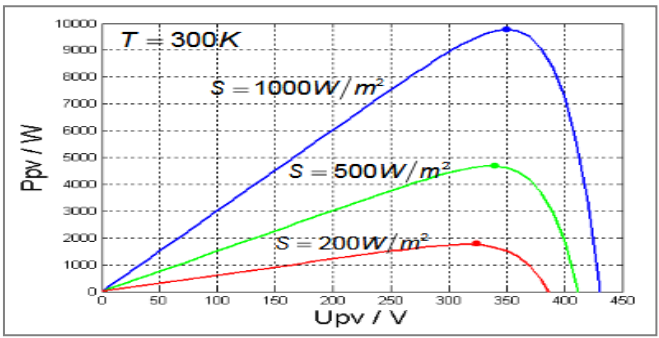

Fig. 14. P vs. as irradiance changes at T=300K

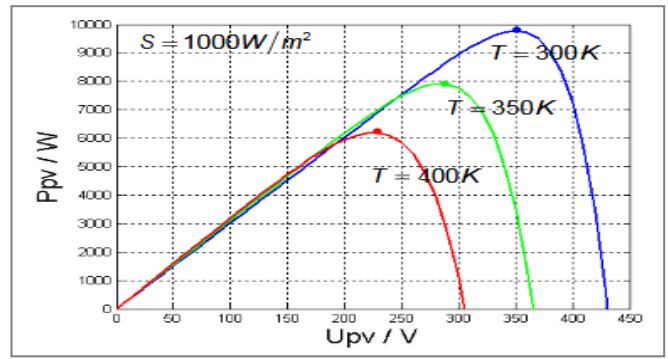

Fig. 15. P vs. V as temperature changes at irradiance (S)=1000W/m2

*B. Maximum Power Point (MPP)*

The power produced by the cell in Watts can be easily calculated along the I-V sweep by the equation P=IV. At the $I_{SC}$ and $V_{OC}$ points, the power will be zero and the maximum value for power will occur between the two. The voltage and current at this maximum power point are denoted as $V_{MP}$ and $I_{MP}$ respectively (Fig. 16) [3].

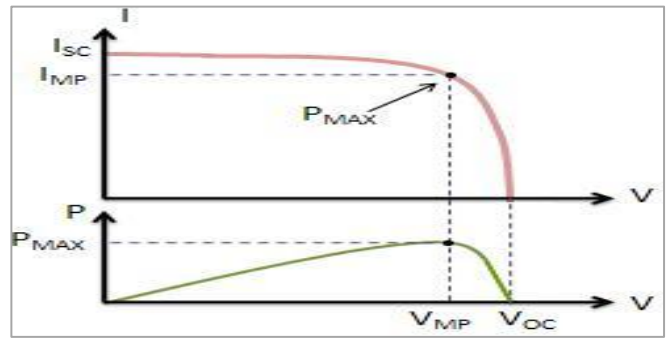

Fig. 16. Maximum power for an I-V sweep

A major characteristic of the PV solar panel is that the available maximum power is provided only in a single operating point given by a localized voltage and current known, called Maximum Power Point (MPP). But the position of this point is not fixed. It moves according to the power per unit area received from the Sun (irradiance) level, the temperature and the load, and the environment (wind .. rain .. etc.) Tracking this point and stabilizing it is very important, and it is called Maximum Power Point Tracking (MPPT) [6].

*C. Maximum Power Point Tracking (MPPT)*

MPPT is basically a load matching problem. A DC to DC converter is required to change the input resistance of the panel to match the load resistance (by varying the duty cycle). It has been observed that the efficiency of the DC to DC converter is maximum for a buck converter, than for a buck-boost converter and minimum for a boost converter [7]. The MPPT working principle is based on the maximum power transfer theory. The power delivered from the source to the load is maximized when the input resistance seen by the source matches the source resistance. Therefore, in order to transfer maximum power from the panel to the load the internal resistance of the panel has to match the resistance seen by the PV panel. For a fixed load, the equivalent resistance seen by the panel can be adjusted by changing the power converter duty cycle. [7]

Controlling the MPPT is usually done by conventional methods such as Pertubate and Observe (P&O) method, which is an iterative method measures a PV module current and voltage, then perturbs the operating point of a PV module to determine the change direction. Fig. 17 shows the flow chart of the classical P&O algorithm [11].

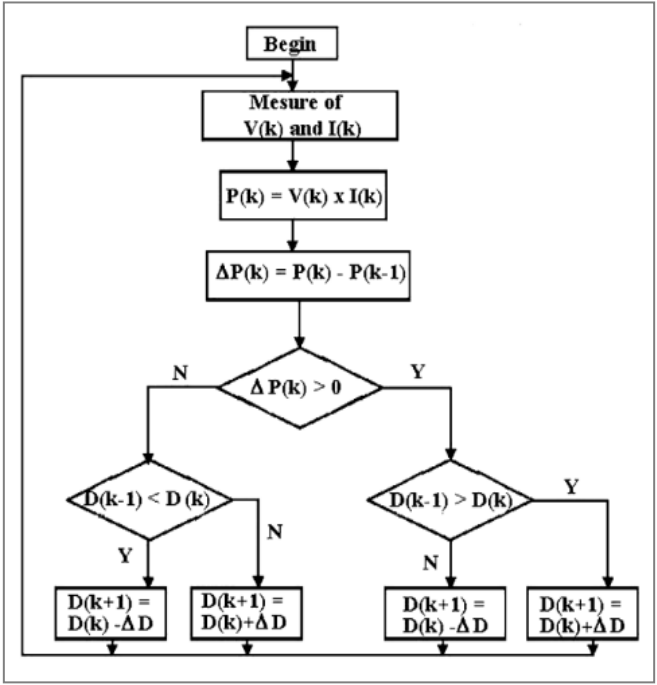

Fig. 17. Flow chart of the classical P&O algorithm

The MPP tracker operates by periodically incrementing or decrementing the solar panel voltage, current or the duty cycle comparing to the PV output power with that of the previous perturbation cycle. If a given perturbation leads to increase (or decrease) the output power of the PV, the successive perturbation is generated in the same (or opposite) direction. On Fig. 18, if the operating point is on the left of MPP (point A), the duty cycle must be decreased until the MPP is reached. If the operating point is on the right of the MPP (point B), the duty cycle is increased to reach the MPP (point C) [12].

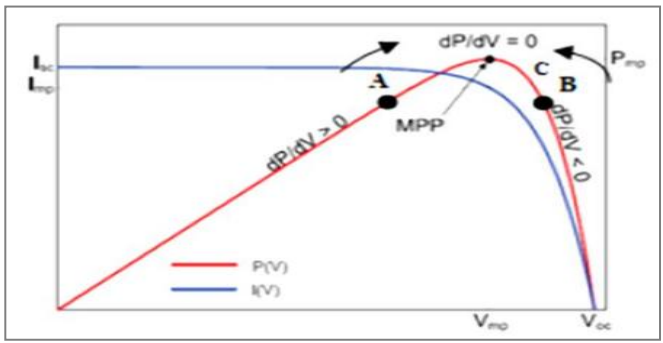

Fig. 18. Sign of dP/dV at different points on the P vs. V curve

Tracking the maximum power point is usually taking advantage of the sensitivity of output voltage to the changes in the duty cycle of the DC-DC converters, which are adapters controlling the load power through a regulated duty cycle (D). In order to step up the voltage, the operation consists of switching an IGBT (An insulated-gate bipolar transistor) (Fig. 19) at a high commutation frequency, with output voltage control by varying the switching duty cycle (D)[15].

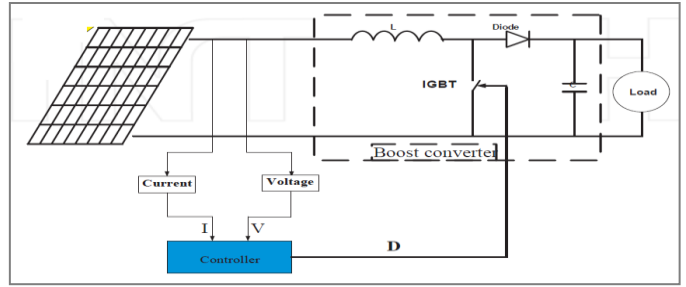

Fig. 19. Changing duty cycle through switching

### D. PV Efficiency (η)

The next step after tracking the maximum power point get it fixed, is improving the PV Efficiency, which is the ratio of the electrical power output $P_{out}$, compared to the solar power input, $P_{in}$, into the PV cell. We must notice that $P_{in}$ which we are talking about here is not the standard $P_{in}$ (power at room temperature and standard irradiance) but the real input power, which depends on the ambient conditions (temperature, shade, irradiance… etc.), while $P_{out}$ can be taken to be $P_{MAX}$ out of the whole system, which includes the PV, regulators, dc-dc, dc-ac, controller type, load, ..etc., according to (5) as follows:

$$\eta = \frac{P_{out}}{P_{in}} => \eta_{MAX} = \frac{P_{MAX}}{P_{in}} \qquad (5)$$

The maximum efficiency ($\eta_{MAX}$) found from a light test is not only an indication of the performance of the device under test, but, like all of the I-V parameters, can also be affected by ambient conditions such as temperature, intensity, spectrum of the incident light, shade..etc. So controlling and improving the PV system efficiency involves many factors and old conventional control methods could not be the efficient way to PV systems improvement, and that is why we decided to use some unconventional, intelligent control techniques to tackle the problem, which is the subject of the following section.

## IV. USING INTELLIGENT CONTROL TO IMPROVE PV SYSTEMS EFFICIENCY

In using intelligent control, namely FLC, we thought it would be a straight forward process, since there were standard parameters that researchers established when dealing with the problem of improving the efficiency of PV systems, but what we found was that important parameters were ignored in the process or assumed to be not as important as the one used in both conventional and intelligent solutions of the problem.

One main point is that outdoor measurements under real conditions are totally different than indoor simulations, and this is especially important when PV systems are used in Libya while relying on simulated cell performance in different continents under different ambient conditions, such as temperature, shade, irradiance, etc.

Working on the practical part of this study, we started with a comparison between the manufacturers specifications and the real measured specifications in the field. The following are the PV-Module Specifications of the PV module that we used in our work, according to the manufacturer (Fig. 20), followed by the real values that we measured in the field (Fig. 21). The differences in values are due to the differences between the environment in the manufacturer's location and ours (The city of Benghazi, Libya).

*A. PV Module Used in Our Application*

We used a PV module that has the specifications showed on Fig. 20 (*Solar Module Type: SP50-36A*):

| Out Peak Power ($P_m$) | 50W | Power Allowable Range: | ±5% |
|---|---|---|---|
| Open Circuit Voltage ($V_{oc}$): | 21.6V | Max Power Voltage ($V_{mp}$): | 18.0V |
| Short Circuit Current ($I_{sc}$): | 3.05A | Max Power Current ($I_{mp}$): | 2.77A |
| Number o cells: | 9x4=36 | Max System Voltage: | 715 VDC |

Measured at standard test conditions (STC):
AM1.5, Irradiance = 1000W/m², T = 25°C
Dimensions: length x width x depth (630 mm x 541 x 35mm)

Fig. 20. PV-Module Specifications according to the manufacturer (SP50-36A)

*B. The Actual PV-Module Measurements (at 25°C)*

We found out that the actual specifications (measured) are different than those specified by the manufacturer, as shown on Fig. 21 below:

| Out Peak Power ($P_m$) | 53.375W |
|---|---|
| Open Circuit Voltage ($V_{oc}$): | 17.5V |
| Short Circuit Current ($I_{sc}$): | 3.05A |
| Irradiance (r) | 403 W/m² |

Fig. 21. Measured PV-Module Specifications in the field (Benghazi, Libya)

*C. PV Parameters Knowledge Base*

Based on our findings in the previous sections; the assumed STC irradiance = 1000 W/m² at 25°C, and measuring temperature correspond to about 1000 W/m², we found it to be 30°C, and the power rose to 62 watts (compare that to the STC max of 55W (50±5%). So, we made enough measurements to get a real practical database, which we used in our application as a knowledge base for FLC; and it is composed of three files (irradiance effect (re), temperature effect (te) and shading effect (se)), noting that most research considers two parameters (in a two dimensional FLC rules map) and our new contribution is to implement three parameters (in a three dimensional FLC rules map).

The knowledge base files are each containing three to five records, and every record composed of four fields, which are: irradiance (r) or temperature (t) or shading (s), open circuit voltage ($V_{oc}$), sort current current ($I_{sc}$) and maximum ($P_{mc}$)(we call it $P_{mc} = V_{oc}$ x $I_{sc}$, to differentiate it from $P_m$, which is $V_m$ x $I_m$)(Fig. 22). Fig. 22 shows the re effect, while Fig. 23 shows the te effect and Fig. 24 shows the se effect. The column on the far right is the fuzzified (linguistic) equivalent of the real variable r, t, or s.

| Pmc | Isc | Voc | Real r | Fuzzy r |
|---|---|---|---|---|
| 2.77 W | 0.15 A | 18.50 V | 403 W/m² | very light |
| 25.61W | 1.30 A | 19.70 V | 650 W/m² | light |
| 61.50 W | 3.00 A | 20.50 V | 1050 W/m² | normal |
| 62.52 W | 3.05 A | 20.50 V | 1060 W/m² | strong |
| 62.73 W | 3.06 A | 20.50 V | 1065 W/m² | very strong |

Fig. 22. re effect on Voc, Isc and Pm

| Pmc | Isc | Voc | Real r | Real t | Fuzzy t |
|---|---|---|---|---|---|
| 53.375 W | 3.05 A | 17.5 V | 403 W/m² | 25 °C | low |
| 60.085 W | 3.05 A | 19.7 V | 605 W/m² | 29 °C | normal |
| 62.730 W | 3.06 A | 20.5 V | 1065 W/m² | 31 °C | high |
| 62.730 W | 3.06 A | 20.5 V | 1420 W/m² | 36 °C | very high |

Fig. 23. te effect on r, Voc, Isc and Pm

| Pmc | Isc | Voc | Real s | Fuzzy s |
|---|---|---|---|---|
| 0.70 W | 0.04 A | 19.6 V | 1420 W/m² | 0.25 |
| 0.15 W | 0.01 A | 15.0 V | 1400 W/m² | 0.50 |
| 0.78 W | 0.40 A | 17.5 V | 1065 W/m² | 1.00 |

Fig. 24. se effect on r, Voc, Isc and Pm

Next we used the same measurements technique while connecting various loads to the PV system and on doing that we concentrated on the third parameter (shading) which is usually ignored by researchers, to see how it affects the MPP (notice here MPP corresponds to ($P_m = I_m$ x $V_m$) which is not on line with ($P_{mc} = I_{sc}$ x $V_{oc}$) (Fig. 25)[13].

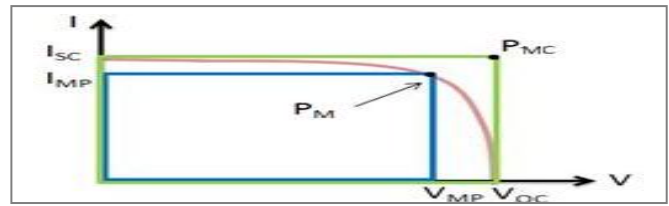

Fig. 25. Location of Pm compared with $P_{mc}=I_{sc}xV_{sc}$

Fig. 26 shows load effect on system's performance under shading and under no shading conditions. The measured values are: load voltage ($V_L$), load current ($I_L$) and load power ($P_L$) at various loads.

| Full Shading | | | No Shading | | | |
|---|---|---|---|---|---|---|
| Power ($P_L$)mW | Current ($I_L$) mA | Voltage ($V_L$) V | Power ($P_L$) mW | Current ($I_L$) mA | Voltage ($V_L$) | Load (L) Ω |
| 397.88 | 20.3 | 19.60 | 408.715 | 19.0 | 21.50 V | 10000 |
| 197.96 | 10.1 | 19.60 | 207.90 | 9.67 | 21.50 V | 9000 |
| 131.32 | 6.70 | 19.60 | 136.31 | 6.34 | 21.50 V | 8000 |
| 98.00 | 5.00 | 19.60 | 102.34 | 4.76 | 21.50 V | 7000 |
| 78.40 | 4.00 | 19.60 | 82.34 | 3.83 | 21.50 V | 6000 |
| 64.68 | 3.30 | 19.60 | 68.80 | 3.20 | 21.50 V | 5000 |
| 54.88 | 2.80 | 19.60 | 58.91 | 2.74 | 21.50 V | 4000 |
| 47.04 | 2.40 | 19.60 | 51.60 | 2.40 | 21.50 V | 3000 |
| 43.12 | 2.20 | 19.60 | 45.79 | 2.13 | 21.50 V | 2000 |
| 37.24 | 1.90 | 19.60 | 41.28 | 1.92 | 21.50 V | 1000 |
| 13.72 | 0.70 | 19.60 | 16.34 | 0.76 | 21.50 V | 0 Ω |

Fig. 26. Load effect on VL, IL and PL at Shading and no shading

Graphically, Fig 27 through Fig. 30 show the relation between $V_L$, $I_L$ and $P_L$, and it appears that the maximum power is affected by changing load values. If load affects MPP, why is it that most researchers concentrate on other parameters, like voltage, current, temperature, irradiance, etc., and not paying attention to this important factor. This point is another contribution of this study to the PV system's efficiency improvement issue.

Many other studies stopped at assuming that the maximum power would be there once the load matches the output impedance of the whole PV system and not considering other loads which would benefit of a maximum power point value just as the matching load, or loads, when thinking dynamic effects of loads on the MPP.

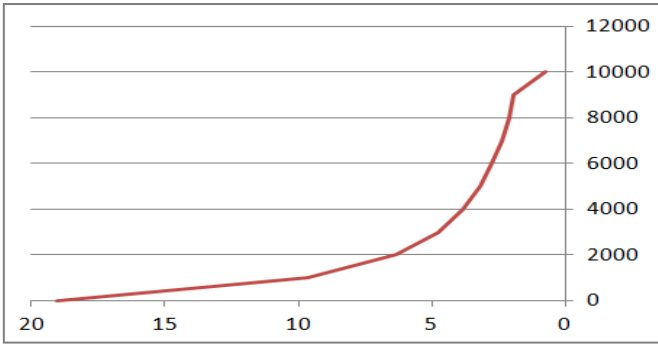

Fig. 27. Current vs. load (no shade)

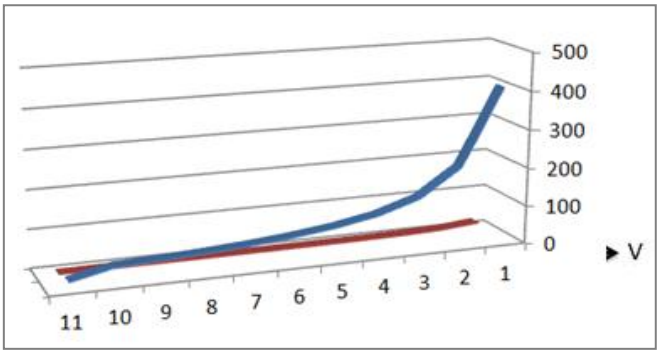

Fig. 28. Power vs. voltage/current (no shade)

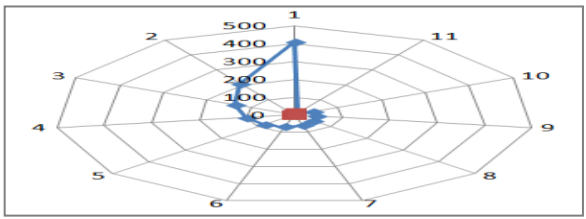

Fig. 29. Power vs. voltage (changing Load)

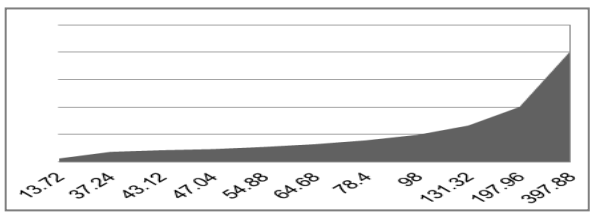

Fig. 30. Power vs. current (partial shading)

What happens when loads were small is shown on Fig. 31 and Fig. 32, which suggests the existence of MPP and its vertical location has little dependence on the shading factor, because negative change in current occurs at positive change in load.

| Power ($P_L$) | Current ($I_L$) | Voltage ($V_L$) | Load (L) |
|---|---|---|---|
| 06.15 | 0.30 | 20.50 | 18 |
| 07.60 | 0.40 | 19.00 | 17 |
| 09.30 | 0.50 | 18.60 | 16 |
| 13.27 | 0.75 | 17.70 | 14 |
| 13.20 | 0.80 | 16.50 | 14 |
| 15.70 | 1.00 | 15.70 | 12 |
| 19.60 | 1.40 | 14.00 | 11 |
| 21.92 | 1.60 | 13.70 | 11 |
| 20.40 | 1.70 | 12.00 | 11 |
| 21.27 | 1.85 | 11.50 | 10 |
| 20.90 | 1.90 | 11.00 | 1.2 |
| 23.32 | 2.20 | 10.60 | 0.5 |
| 25.50 | 2.50 | 10.20 | 18 |
| 03.60 | 3.00 | 01.20 | 17 |
| 01.70 | 3.10 | 0.550 | 16 |

Fig. 31. Load effect on VL, IL and PL

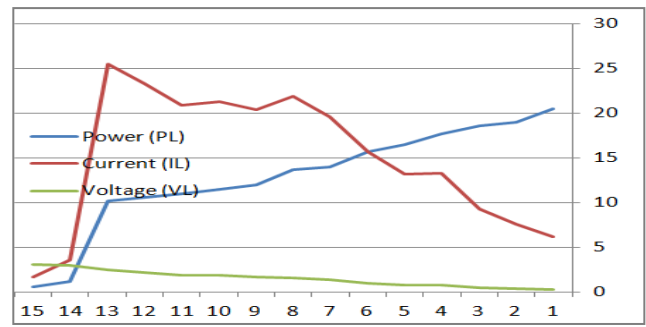

Fig. 32. Load effect, p, v and i

*D. FLC with Probability Feedback*

Notice that in general making measurements while fixing some parameters might be an appealing idea, but as we could see from studying the data above, practical work shows that parameters are dynamically changing and uncertainty arise, so we found it helpful to borrow a concept called "Explicit Congestion Notification (ECN)" from telecommunications congestion control, and utilized in our FLC design. The idea is to use the probability of congestion other than congestion itself and in our case we used the probability that the power point is at the MPPT other than being at the MPPT itself, that helps in predicting where the power point would be so actuators or soft computing could change parameters dynamically and in accordance with the probability, so Instead of using the maximum power directly to change the direction (increase or decrease) we used the probability of ($p_{max}$, less than and more than $p_{max}$), if $p=p_{max}$ no action needed, and if $p<p_{max}$ then we need increase it, and if $p>p_{max}$ then power must be decreased.

Fig. 33 shows a sample of our parameter values and how it is classified according to the probability (as in Fig. 44) that it is marked for "no change in power: at maximum power point (mpp)", "increase power at different levels: vhi=very high increase, hi=high increase, mi=medium increase, li=low increase" and "decrease power at different levels: ld=low decrease, md=medium decrease, hd=high decrease, vhd=very high decrease". The actions could be implemented by the controller actuators and or by its soft computing options

| Change in Im | Change in Vm | Change in Pm | How far Pm from MPP | Probability (collapsed) | action |
|---|---|---|---|---|---|
| -0.0445 | 0.4 | -0.178 | very far down | 1.0 | vhi |
| -0.040 | 0.3 | -0.120 | far down | 0.9 | hi |
| -0.042 | 0.2 | -0.084 | near down | 0.8 | mi |
| -0.044 | 0.1 | -0.040 | very near down | 0.7 | li |
| 0 | 0 | 0 | mpp | 0.6 | na |
| 0 | 0 | 0 | mpp | 0.5 | na |
| +0.043 | 0.1 | +0.043 | very near up | 0.4 | ld |
| +0.040 | 0.2 | +0.080 | near up | 0.3 | md |
| +0.041 | 0.3 | +0.123 | far up | 0.2 | hd |
| +0.043 | 0.4 | +0.172 | Very far up | 0.1 | vhd |

Fig. 33. Fuzzy logic rules / probability / action

The result is a crisp value at the output of the fuzzy logic control system, which is a feedback FLC (Fig. 34). This is another contribution of our study that might have been expressed before but not as clear as ours.

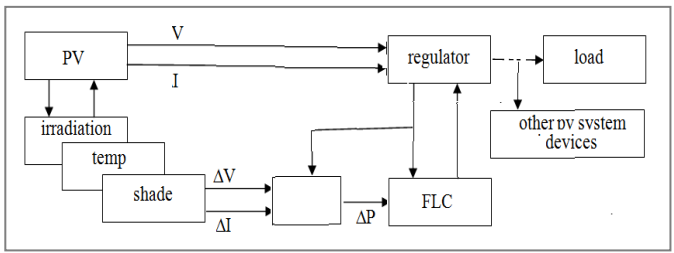

Fig. 34. FLC system as part of the whole PV system

In standard FLC two parameters change, usually the error and the change in power with respect to change in voltage yields (6) and (7) as follows:

$$E(k) = \frac{P(k) - P(k-1)}{V(k) - V(k-1)} \qquad (6)$$

$$CE(k) = E(k) - E(k-1) \qquad (7)$$

But, instead of this standard 2D method of FLC where two variables are chosen like error E(k) and change of error CE(k) in (6) and (7), we used a 3D FLC rules table, a sample of which is shown on Fig. 35, it is done using MS Excel, and it is still under study to be expanded to more than three dimensional FLC rule tables.

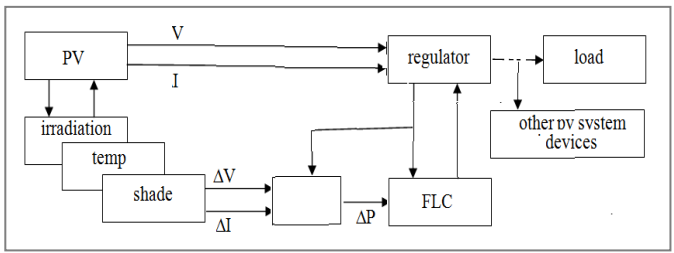

Fig. 35. A sample of a 3D FLC rules table

The third parameter would be either the shading change, the load change, shunt resistance (which represent the loss of current due to ambient conditions), series resistance (which represent the loss of voltage due to ambient conditions), and/or all of these parameters together could change the power curve and its instantaneous values drastically (Fig. 36) [13]. It would be very interesting to use more than 3D, but as we complicate the FLC rules and increase its number we cause slow processing speed.

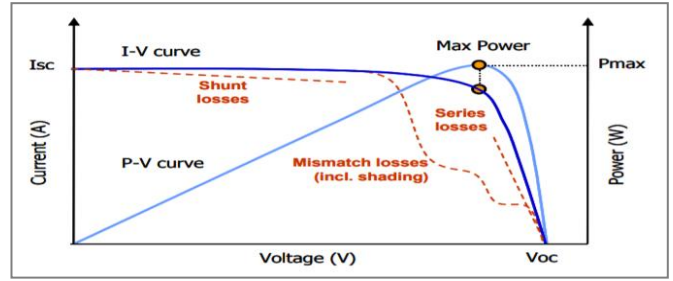

Fig. 36. Ambient conditions affect on the I-V-P curves

### E. FLC Member Functions

Figures (Fig. 37 through Fig. 39) show the input member functions of the FLC system: input1: how near to the MPP, input2: change in v, input3: change in p.

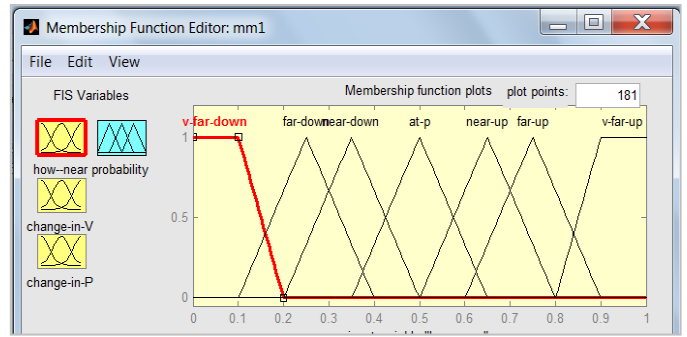

Fig. 37. Input1 member function: how near from mpp

Fig. 4.40 shows output member function, output: probability. Fig. 4.41 shows the FLC rules and Fig. 42 shows the rule viewer of a sample run of the FLC, while Fig. 43 shows it in a three dimension view.

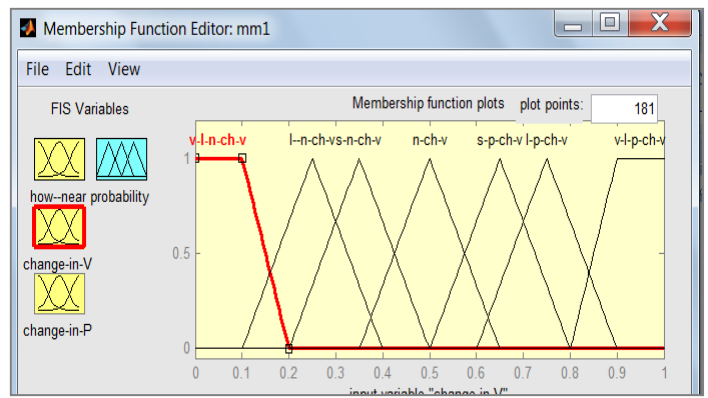

Fig. 38. Input2 member function: change in v

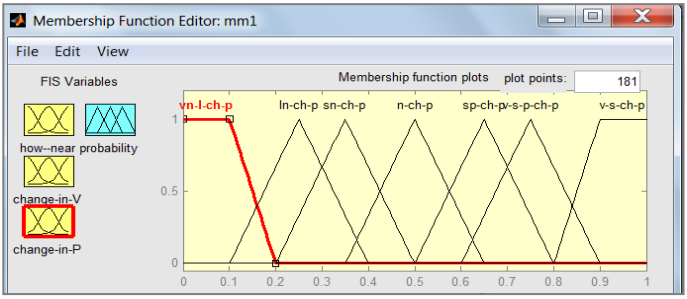

Fig. 39. Input3 member function: change in p

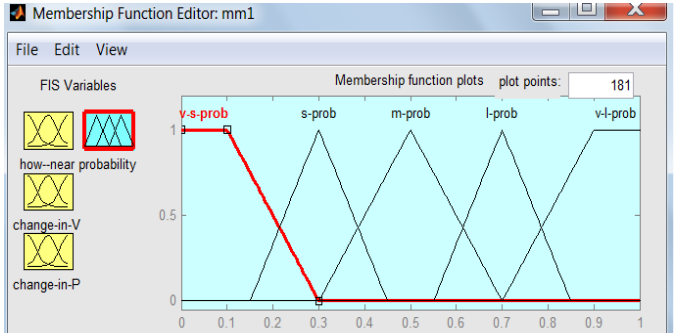

Fig. 40. Output member function: probability

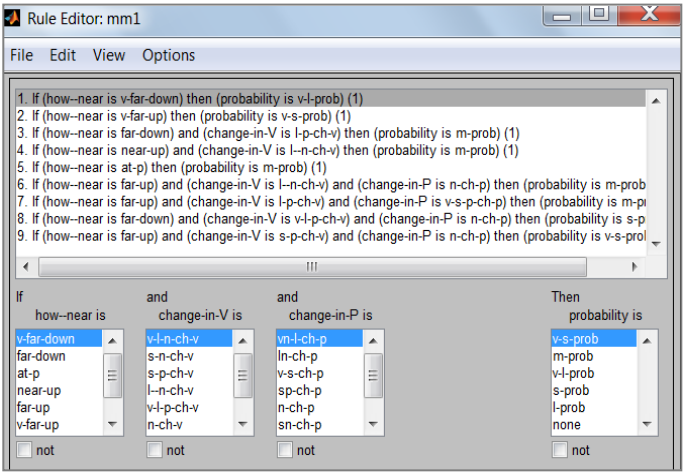

Fig 41. FLC rules

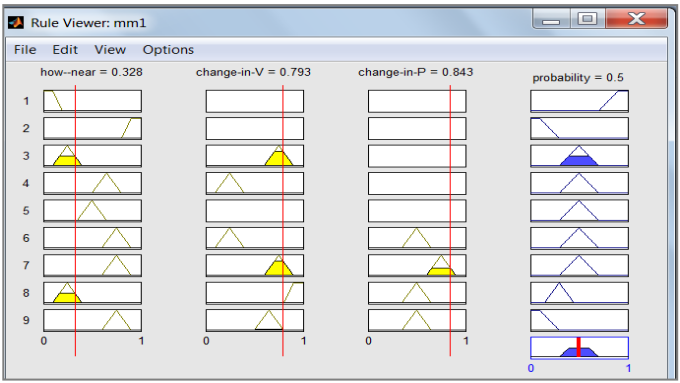

Fig. 42. a sample run of the rule viewer

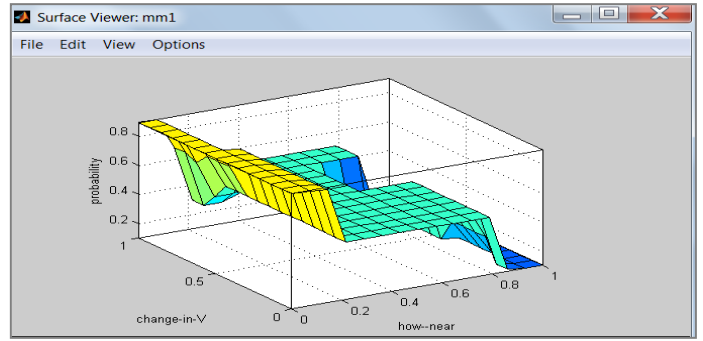

Fig. 43. A sample run in a three dimensional view

Using the Fuzzy Inference System file (mm1.fis), we fed the probability to the controller so the actuator may change parameters that would satisfy Fig. 33, which is an extension to the FLC rules. For example, probability of 0.9 leads to high increase (hi) in resistance (this parameter could be changed by the user as desired). we chose the resistance as a parameter according to (8) and (9) [12], which gives the optimal resistance that leads to optimal efficiency:

$$R_{opt} = P_m/(I_m)^2 \;,\; R_{opt} = 7.9 \times 10^{-5} G^2 - 0.12G + 49 \quad (8)$$

where G is irradiance. And $R_{opt}$ is the optimal value of resistance that leads to optimal efficiency: $\eta = I_m \times V_m / G \times A$, and generally resistance is:

$$R = \frac{\ln(\frac{Isc-Im}{Il}) n \times VT - Vm}{Im} \quad (9)$$

where n = number of cells, Il = load current and $V_T$ = nkT/q.

Theoretically, it was found that the resistance is 8Ω, and in our case it was found to be 11Ω, which is very close, and the small difference is due different cell specifications used that we mentioned at the beginning of this chapter.

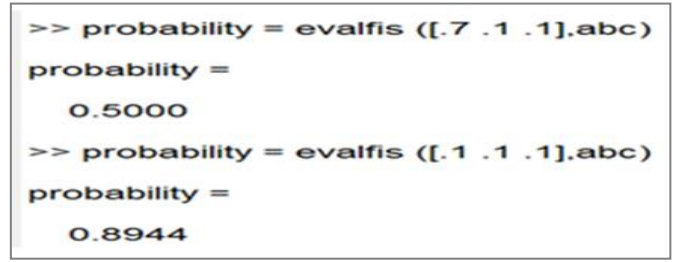

Fig. 44. FLC evaluation at a sample inputs

In working on the improvement of the efficiency, we avoided using the standard ways of tackling such problem; we used probability to predict how far the power from the MPP and let the controller act before not after the power got away from the MPP either down or up. The other thing we made our measurements as more parameters than usual change; and that showed surprisingly strange looking power and voltage curves than the ones we usual see in simulations. In the following section we list some concluding remarks and some recommendations that could improve PV systems efficiency.

## V. Conclusions

Although PV systems are the cleanest and the easiest to implement, a lot of work needed to increase its efficiency, since most of the work done concentrated on the usual standards and conventional classic control techniques and also relied on simulations rather than field work. In this work we tried to tackle the issue of improving PV by relying on the field work beside the academic research that we studied throughout doing this work. we considered many parameters rather than the two parameter standard used in most research, and we implemented probability rather than certainty to predict the MPP and take actions before not after the MPP move up or down. we also used three dimensional FLC with feedback which seems to work best with microcontrollers and soft computing. We recommend using more than one intelligent technique although that might lead to slow processing and we also think using quantum probability rather than regular probability, which means the measured probability (the collapsed value) is a result of many probable values at the same time. Another thing that could be investigated is the chaotic behavior of the PV system, i.e. little changes in parameters could lead to drastic change in the output, that would mean getting huge amount of energy out of little amount of available resources.

## References

1. Renewables 2016 Global Status Report, Ren21 Secretariat, Paris 2016 www.ren21.net/wp-content/uploads/2016/05/GSR_2016_Full_Report_lowres.pdf
2. Gil Knier, How do Photovoltaics Work?, NASA https://science.nasa.gov/science-news/science-at-nasa/2002/solarcells
3. Photovoltaic Cell I-V Characterization Theory, National Instruments, 2016  http://www.ni.com/white-paper/7230/en/
4. Bax Smith , Classical vs Intelligent Control,Special Topics in Robotics, 2002.  https://www.engr.mun.ca/~baxter/Publications/ClassicalvsIntelligentControl.pdf
5. Fuzzy Logic, Wikimedia, 2017  https://en.wikipedia.org/wiki/Fuzzy_logic
6. A. Mellit, et al., AI Techniques for Sizing Photovoltaic Systems, Elsevier, 2017. http://www.sciencedirect.com/science/article/pii/S1364032108000051
7. B. Amrouche et al.,Artificial intelligence based P&O MPPT, ICRESD-07 Tlemcen, 2007. http://www.cder.dz/download/ICRESD07_3.pdf
8. Farzad Sedaghati et al., PV MPPT by Using ANN, Mathematical Problems in Engineering, 2012. www.hindawi.com/journals/mpe/2012/506709
9. A. Kulaksiz and O. Aydoğdu, ANN-Based MPPT of PV Systems Using Fuzzy  Controller, http://ieeexplore.ieee.org/document/6246936/
10. M. Makhloufi et al.,Tracking Power Photovoltaic System using ANN Control Strategy, .J. Intelligent Systems and Applications, 2014 http://www.mecs-press.org/ijisa/ijisa-v6-n12/IJISA-V6-N12-3.pdf
11. Po-Chen Cheng et al., Optimization of a Fuzzy-Logic-Control-Based MPPT Algorithm Using the Particle Swarm Optimization, Energies 2015 http://www.mdpi.com/1996-1073/8/6/5338
12  Nerissa et al. A Model for PV Module Optimization, Journal of Mechanical Engineering and Automation, 2015. http://www.sapub.org/global/showpaperpdf.aspx?doi=10.5923/j.jmea.20150502.02
13. Guide To Interpreting I-V Curve Measurements of PV Arrays, Solmetric Corp, 2010. http://resources.solmetric.com/get/Guide to Interpreting  I-V Curves.pdf